\pdfoutput=1
\documentclass[a4paper,american,floatfix,pdftex,superscriptaddress,twoside,%
aip,jcp,%
reprint,
]{revtex4-2}%
\usepackage[T1]{fontenc}
\usepackage{multirow}
\usepackage{graphicx}%
\usepackage[utf8]{inputenc}
\usepackage{hyperref, hypernat}
\usepackage[ams,todos]{CMImacros}
\usepackage[boxed]{algorithm2e} 

\usepackage{tabularx}
\usepackage{longtable}
\usepackage{booktabs}
\newcommand{\cfeldesy}{\affiliation{Center for Free-Electron Laser Science CFEL, Deutsches
      Elektronen-Synchrotron DESY, Notkestr. 85, 22607 Hamburg, Germany}}%
\newcommand{\uhhphys}{\affiliation{Department of Physics, Universität Hamburg, Luruper Chaussee 149,
      22761 Hamburg, Germany}}%
\newcommand{\uhhmaths}{\affiliation{Department of Mathematics, Universität Hamburg, Bundesstr. 55,
      20146, Hamburg, Germany}}%
\newcommand{\theochem}{\affiliation{Institute for Theoretical Chemistry, University of Stuttgart, Pfaffenwaldring 55, 70569 Stuttgart, Germany}}%
\newcommand{\evemail}{\email[Email:~]{emil.vogt@cfel.de}}%
\newcommand{\ayemail}{\email[Email:~]{andrey.yachmenev@theochem.uni-stuttgart.de}}%
\newcommand{\cmiweb}{\homepage[\\ URL:~]{https://github.com/robochimps}}%

\SetAlCapSkip{\abovecaptionskip}

\begin{document}
\title{Transferability and interpretability of vibrational normalizing-flow coordinates}%
\author{Emil Vogt}\evemail\cfeldesy %
\author{Álvaro Fernández Corral}\cfeldesy\uhhphys %
\author{Yahya Saleh}\cfeldesy\uhhmaths %
\author{Andrey Yachmenev}\ayemail\cmiweb\theochem%
\date{\today}

\begin{abstract}
The choice of vibrational coordinates is crucial for the accuracy, efficiency, and interpretability of molecular vibrational dynamics and spectra calculations.
We explore the recently proposed \emph{normalizing-flow} vibrational coordinates, which are learned molecule-specific coordinate transformations optimized for a given basis set.
Much like how spherical coordinates naturally simplify the hydrogen atom by embedding physical insight into the coordinate system, normalizing-flow coordinates offload complexity from the basis functions into the coordinate transformation itself. This shift not only improves basis-set convergence, but also enhances interpretability of vibrational motions. We provide an analysis of the utility, interpretation and associated constraints of normalizing-flow vibrational coordinates.
Moreover, we demonstrate that these coordinates can be generalized across different isotopologues, and even structurally related molecules, achieved with minimal fine-tuning of selected output parameters.
\end{abstract}

\maketitle

\section{Introduction}
Theoretical and computational studies of molecular vibrations are central to
theoretical chemistry, molecular physics, and related scientific fields. Key
areas of interest include the calculation and fitting of accurate potential
energy surfaces
(PESs)~\cite{Braams:IRPC28:577,Manzhos:CR121:10187,Mizus:PTRSA376:20170149, Saleh:JCP155:144109},
the development of effective molecular Hamiltonians for efficient
representations of molecular
spectra~\cite{Bowman:IRPC22:533,Franke:JPCA125:1301,Hansen:JCTC6:235}, and the
first-principles computation of accurate vibrational energies, wavefunctions,
and spectra~\cite{Yurchenko:JMS245:126,  Matyus:JCP130:134112,
Bowman:IRPC22:533, Hansen:JCTC6:235}. The latter can be accomplished with
perturbation theory methods, variational theory approaches using finite-basis
expansions, or pseudo-variational methods like discrete-variable
representations. Central to all these methods is the construction of an
appropriate vibrational Hamiltonian, which relies on carefully chosen
coordinates and associated basis functions. The selection of coordinates plays
an important role in defining the Hamiltonian operator, influencing the extent
to which vibrational motions are
coupled~\cite{Bramley:MP73:1183,Mendolicchio:JCTC18:7603} and thus the overall computational efficiency. Determining an
optimal coordinate system and compatible basis functions for describing
molecular vibrations often requires substantial expertise and prior knowledge of
the vibrational motions.

Rectilinear normal coordinates are effective for computing low-energy states in
semi-rigid molecules, where the PES often can be relatively well approximated by a low-order
Taylor-series expansion around a single equilibrium geometry. However, these
coordinates become inadequate in calculations of delocalized states, such as
those encountered in floppy molecules~\cite{Matyus:ChemComm59:366,
Bacic:ARPC40:469}, or high-energy states in semi-rigid
molecules~\cite{Simko:ACIE62:e202306744,Yachmenev:JCP143:014105}, which sample
larger and more complex regions of the phase space.

Curvilinear coordinates like Radau~\cite{Wang:JCP129:234102}, Jacobi~\cite{Gatti:JCP108:8804, Leforestier:JCP114:2099}, valence~\cite{Bramley:JCP98:1378}, ellipsoidal~\cite{Bacic:JCP90:3606}, hyperspherical~\cite{Chapuisat:MP72:1}, and polyspherical~\cite{Iung:IJQC106:130, Gatti:PhysRep484:1, Klinting:JCTC16:4505} types are often better suited for vibrational calculations in floppy molecules~\cite{Bowman:MP106:2145, Matyus:JCP130:134112}. The optimal choice depends on the specific nuclear motions involved and requires an understanding of the morphology of the PES. Because vibrational motions vary widely across different molecular structures, no single coordinate system is universally optimal. The general approach for selecting effective coordinates is to capture the primary variations in the PES along individual dimensions, which helps to minimize coupling between different vibrational modes. Chemical intuition suggests that valence coordinates are most suitable for many molecules, as the localized electron density between adjacent atoms leads to significant changes in the PES with variations in bond lengths.

A promising strategy is to use general parametrized coordinates, with parameters optimized in variational calculations of vibrational energies. Optimized linear combinations of normal or valence coordinates have been explored in several studies~\cite{Thompson:JCP77:3031, Mayrhofer:TCA92:107,Oenen:JCP160:014104, MendiveTapia:JCTC19:1144, Schneider:JCP161:094102,Zuniga:JCP122:224319,Bowman:JCP90:2708, Yagi:JCP137:204118, Thomsen:JCP140:154102,Bulik:JCP147:044110}. While this strategy has demonstrated an improvement in basis-set convergence of energy calculations, it has not gained widespread adoption.
One noteable limitation is that the coordinate transformation often destroys the truncated form of the PES, e.g., an $N$-mode expansion, constructed in the original coordinates.
This leads to higher coupling orders between coordinates and, consequently, more expensive evaluations of vibrational integrals in sum-of-products formulations of the PES and kinetic energy operators. In addition, the improvements achieved with optimized linear coordinate mappings is often modest, due to their restrictive nature. General coordinate transformations also tend to break symmetry, which can be important for maintaining computational efficiency~\cite{Ziegler:JCTC15:4187}.

Once the vibrational coordinates have been selected, the next step is to choose
an appropriate basis set, guided by the domain of each coordinate. For example,
valence coordinates have domains of $[0,\infty)$ for bond lengths, $[0,\pi]$ for
angles, and $[0,2\pi]$ for dihedrals. A common approach is to use a
direct-product basis of univariate orthogonal polynomial-based functions (or linear combinations thereof), primarily due to their close relation with
Gaussian quadratures~\cite{Bacic:ARPC40:469,Wang:JCP130:094101}. This approach
simplifies the evaluation of integrals necessary for the calculation of the
Hamiltonian matrix elements and facilitates transformations between the
finite-basis representations and their corresponding discrete-variable
representations. The specific univariate functions are chosen based on the
coordinate domains, the shape of the PES along the coordinates, and the degree of
vibrational coupling in the Hamiltonian. 

We recently introduced a new approach for nonlinear parametrization of
vibrational coordinates based on normalizing flows~\cite{Saleh:JCTC21:5221,Papamakarios:JMLR22:1}, implemented
through an invertible residual neural network~\cite{Behrmann:ICML2019:573}. In the machine
learning literature, normalizing flows refer to a sequence of invertible and
differentiable transformations~\cite{Rezende:ICML37:1530, Papamakarios:JMLR22:1}. In our method, the neural network
parameters are optimized using the variational principle to minimize the
approximate energies. The coordinates are thereby optimized to maximize the
performance of a chosen truncated basis set, tailoring it to the specific
molecule. We note that several studies have proposed similar frameworks - such as optimized-coordinate Vibrational Self-Consistent Field (VSCF) theory - where the coordinates are also parametrized and optimized variationally to minimize approximate energies~\cite{Thompson:JCP77:3031,Bacic:JCP90:3606,Ziegler:JCTC15:4187,Zuniga:JCP122:224319,Thomsen:JCP140:154102,Bulik:JCP147:044110}. While these approaches can be tailored to the specific molecule to some extent, they typically rely on a fixed functional form for the coordinates, which limits their expressivity. In contrast, our use of invertible neural networks enables a more flexible and expressive construction of coordinates, allowing for finer adaptation to the molecular system at hand.

In our previous work~\cite{Saleh:JCTC21:5221}, normalizing-flow coordinates were applied to molecules such as H$_2$S, H$_2$CO and HCN/CNH. It was demonstrated that the learned coordinates enhance the separability of the Hamiltonian, enabling more reliable assignment of approximate quantum numbers. Furthermore, we demonstrated the transferability of these coordinates across different basis-set truncations. Normalizing-flow coordinates have also been used in conjunction with Monte Carlo integration to tackle larger molecular systems, such as CH$_3$CN and C$_2$H$_4$O~\cite{Zhang:JCP161:024103}, and to investigate anharmonic effects in lithium solids at finite temperatures~\cite{Zhang:arXiv2412.12451}.

In this work, we investigate interpretability and effectiveness of
normalizing-flow coordinates. Specifically, we show that when optimized using
the variational principle, these coordinates shift the average density center
(defined as the trace of the position operator matrix divided by the number of basis functions)
to align with that of the eigenbasis. 
This observation offers a practical and physically meaningful way of
interpreting how coordinate transformations influence the representation of
quantum states. Additionally, we demonstrate that the nonlinear mappings enabled by
normalizing flows offer significant advantages over linear transformations, particularly in
capturing anharmonic behavior across semi-finite and finite domains.

We further investigate the transferability of normalizing-flow vibrational
coordinates. Coordinates trained for H$_2$S are found to generalize well to its
deuterated isotopologues (D$_2$S and HDS) with only minor, physically
motivated adjustments. Remarkably, the same coordinate system also
performs well for H$_2$O, suggesting that the learned
coordinates capture structural motifs common to chemically related molecules. In all cases considered,
normalizing-flow coordinates outperform traditional curvilinear valence-bond
coordinates, as evidenced by faster convergence of variationally computed
vibrational energies.

These findings are significant for two main reasons. First, they suggest that
normalizing flows provide a practical means of transferring optimized coordinate
systems across chemically related systems, thereby improving computational
efficiency and initialization. Second, and more broadly, our method offers a
general computational framework for identifying informative coordinate systems - those
that simplify vibrational complexity and improve spectral convergence. While the
notion of intrinsic coordinates is often invoked in chemical intuition to
describe a minimal set of variables underlying molecular motion, our results
suggest that coordinates optimized for computational efficiency may also reflect
such intrinsic structures. Importantly, the optimality of these
coordinates is relative: it depends on both the chosen truncated basis and the target vibrational states.

To build physical insight and isolate key effects, we focus primarily on illustrative one-dimensional systems. These simplified models serve as a clean testbed for understanding the mechanisms and interpretability of variational coordinate optimization using normalizing flows.

\section{Theory}
In variational basis representations, the vibrational Schrödinger equation,
\begin{align}
	\hat{H}\Psi_n = \left( \hat{T} + \hat{V} \right)\Psi_n = E_n\Psi_n,
\end{align}
is projected onto a finite set of orthonormal basis functions. Here, $\hat{H}$
is the vibrational Hamiltonian operator, $\hat{T}$ is the kinetic energy operator
and $\hat{V}$ is the potential energy operator. $\Psi_n$ and $E_n$ are the $n$-th eigenfunction and eigenvalue,
respectively.
The vibrational eigenfunctions, $\Psi_n$ $(n=0\dots N-1)$, are approximated as linear combinations of $M$ basis functions $\{ \phi\}_{m=0}^{M-1}$, with $M\geq N$, as 
\begin{align}\label{eq:Psi_m}
	\Psi_n({\mathbf r}) \approx \tilde{\Psi}_n({\mathbf r}) = \sum_{m< M} c_{nm}\phi_m({\mathbf r}),
\end{align}
where $\mathbf r$ denotes the vibrational coordinates. 

By introducing this linear expansion into the weak formulation of the Schrödinger equation, one obtains a matrix eigenvalue problem
\begin{align}
	\mathbf{H}\mathbf{C} = \mathbf{S}\mathbf{C}\mathbf{E},
\end{align}
where $\mathbf{H}=\{\langle \phi_m|\hat{H}|\phi_m'\rangle\}_{m,m'=0}^{M-1}$ is the Hamiltonian matrix, $\mathbf{C} = \{c_{mn}\}_{m=0,\ n=0}^{M-1,\ N-1}$ are the linear expansion
coefficients, $\mathbf{S}=\{\langle \phi_m|\phi_m'\rangle\}_{m,m'=0}^{M-1}$ is the overlap matrix, and  $\mathbf{E}=\{E_n\}_{n=0}^{N-1}$ are the approximated
vibrational energies.
The accuracy of calculated energies can be systematically improved by increasing
the number of basis functions $M$, ensuring variational convergence to the true
energies as lower bound.
In practice, quadratures or truncated Taylor-series expansions are often employed to evaluate
the integrals required to construct the Hamiltonian matrix elements, which
introduce additional errors to the truncated basis representation and may result in a violation of the variational principle.

An alternative approach to systematically improve calculated energies is to enhance the approximation power of the chosen basis functions. To achieve this, we start with a truncated set of orthonormal basis functions $\{\phi_m(\mathbf q)\}_{m=0}^{M-1}$ defined on a coordinate set $\mathbf q$. The coordinate set $\mathbf q$ is related to an initial set of vibrational coordinates $\mathbf{r}$ through a parametrized invertible mapping $f_\theta$, such that ${\mathbf{q}} = f_\theta(\mathbf{r})$. Because the mapping is invertible, the reverse relation also holds, $\mathbf{r} = f_\theta^{-1}(\mathbf{q})$.
The augmented basis functions are then defined as
\begin{align}\label{eq:gamma_n}
	\gamma_m({\mathbf{q}}; \theta) = \phi_m(\mathbf{q})\sqrt{D}, 
\end{align} 
where $D = \lvert1/\det (\nabla_{\mathbf{q}} f_\theta^{-1}(\mathbf{q}))\rvert = \lvert\det (\nabla_{\mathbf{r}} f_\theta(\mathbf{r}))\rvert$ is the absolute value of the inverse of the determinant of the Jacobian. The inclusion of the factor $\sqrt{D}$ ensures that the augmented basis functions remain orthonormal, regardless of the values of the parameters $\theta$. These basis functions can also be evaluated in the vibrational coordinates $\mathbf{r}$ as $\gamma_m (f_\theta(\mathbf{r}); \theta)$.  
	
In principle, the mapping \( f_\theta \) can be any differentiable invertible function. However, for the augmented basis set to remain complete, \( f_\theta \) must be bi-Lipschitz~\cite{Saleh:arXiv2406:18613}. This means that there exist constants \( k, K > 0 \) such that \( k \leq  1/|\det (\nabla_{\mathbf q} f^{-1}_\theta({\mathbf q})) | \leq K \) for all \( \mathbf{q} \) within the domain. We represent \( f_\theta \) as a normalizing flow implemented through an invertible residual neural network (iResNet)~\cite{Behrmann:ICML2019:573}, which is bi-Lipschitz by construction.

Matrix elements of the vibrational kinetic and potential energy operators can be expressed within the augmented basis in \eqref{eq:gamma_n}, by applying the coordinate transformation ${\mathbf{q}} = f_\theta(\mathbf{r})$. For the potential energy, this leads to the expression: 
\begin{align}\label{eq:potential} 
\mathbf{V}_{mm'} &= \int \phi_{m}^*(f_\theta(\mathbf{r})) \sqrt{D} \ V		(\mathbf{r}) \phi_{m'}(f_\theta(\mathbf{r})) \sqrt{D} \ \mathrm{d}\mathbf{r}  \\
&= \int \phi_{m}^*(\mathbf{q}) V(f_\theta^{-1}(\mathbf{q})) \phi_{m'}(\mathbf{q}) \ \mathrm{d}\mathbf{q}, \nonumber
\end{align} 
where the volume elements of integration are related by $ \mathrm{d}\mathbf{q} =  D\mathrm{d}\mathbf{r}$.
This formulation shows how the change of coordinates effectively modifies the
operators within matrix elements in the original basis set
$\{\phi_m\}_{m=0}^{M-1}$. Therefore, optimizing $f_\theta$ for enhancing the
expressivity of basis functions in \eqref{eq:gamma_n} is equivalent to
optimizing the coordinates in which the Hamiltonian operator is expressed, for the chosen fixed
set of basis functions. 

The corresponding expression for the kinetic energy matrix elements after the coordinate transformation is
\begin{align}\label{eq:keo}
	&T_{mm'} = \int \phi_m^{*}(f_\theta(\mathbf{r})) \sqrt{D} \ \hat{T}(r)\ \phi_{m'}(f_\theta (\mathbf{r})) \sqrt{D} \ \mathrm{d}\mathbf{r} & \\
	&= \frac{\hbar^2}{2} \sum_{kl} \int \left[ \left(\frac{1}{2\sqrt{D}}\frac{\partial D}{\partial q_k} + \sqrt{D}\frac{\partial}{\partial q_k}\right) \phi_m^*(\mathbf{q})\right]& \nonumber \\
	&\sum_{\lambda\mu}\frac{\partial q_k}{\partial r_\lambda}G_{\lambda\mu}(f_\theta^{-1}(\mathbf{q}))\frac{\partial q_l}{\partial r_\mu}& \nonumber \\
	&\left[ \left(\frac{1}{2\sqrt{D}}\frac{\partial D}{\partial q_l} + \sqrt{D}\frac{\partial}{\partial q_l}\right) \phi_{m'}(\mathbf{q})\right] \mathrm{d}\mathbf{q},& \nonumber
\end{align}
where $G_{\lambda\mu}$ are the elements of the mass-weighted metric tensor (Wilson $G$-matrix). In addition, the pseudo-potential term,
  \begin{align}\label{eq:pseudo}
	&U= 
	\frac{\hbar^2}{32} \sum_{\lambda}\sum_{\mu}
	\frac{G_{\lambda\mu}}{\tilde{g}^2} \frac{\partial\tilde{g}}{\partial r_\lambda}
	\frac{\partial\tilde{g}}{\partial r_\mu}
	+4\frac{\partial}{\partial r_\lambda} 
	\Bigg(
	\frac{G_{\lambda\mu}}{\tilde{g}}
	\frac{\partial\tilde{g}}{\partial r_\mu}
	\Bigg),&
\end{align}
where $\tilde{g} = \det \left( \mathbf{G}^{-1} \right)$,  also contributes to the exact kinetic energy operator. The pseudo-potential arises from the original coordinate transformation from Cartesian to the initial coordinates $\mathbf{r}$. As the pseudo-potential is a scalar function of the vibrational coordinates, its associated matrix elements can be evaluated analogously to the potential energy operator in \eqref{eq:potential}.  

\subsection{Invertible neural networks}\label{sec:nn}
To model the normalizing flow $f_\theta$, we used an iResNet consisting of 10 blocks (five blocks for the one-dimensional examples, \textit{vide infra}). An iResNet is given by concatenating blocks of the form
\begin{align}
 \mathbf{x}_{k+1} = \mathbf{x}_k + \mathbf{h}_k(\mathbf{x}_k; \theta),
\end{align}
where $\mathbf{x}_k$ is the input to the block and $ \mathbf{h}_k(\mathbf{x};
\theta)$ is a feed-forward neural network
composed of
weights, biases and nonlinear activation functions. Each block was constructed as a dense neural network with two hidden layers with unit sizes $[8,8]$, and an output layer with the number of units equal to the number of coordinates. A block is guaranteed
to be invertible if it has a Lipschitz constant < 1. The
inverse of each block was obtained by fixed-point iterations. 

To guarantee that the feed-forward networks $ \mathbf{h}_k$ for $k=0, \dots, K-1$
satisfy the aforementioned Lipschitz condition, we used the LipSwish activation
function
\begin {align}
\sigma(x) := \frac{x}{1.1} \frac{1}{1+\exp(-x)},
\end{align}
which has a Lipschitz constant of $\sim1$. With this activation function, the block $\mathbf{h}_k$ is guaranteed to be Lipschitz if each of its weight
matrices, $\mathbf{W}$, has a spectral norm  $<1$. This is achieved by setting 
\begin {align}
\mathbf{W}= \begin{cases}
	\mathbf{W} &\quad\text{if } \|\mathbf{W}\|_2 < c\\
	\mathbf{U}\tilde{\mathbf{\Sigma}}\mathbf{V}^T &\quad\text{if } \|\mathbf{W}\|_2 \geq c,
\end{cases}
\end{align}
where $0<c<1$ is a hyperparameter, $\|\mathbf{W}\|_2$ is the
spectral norm of the weight matrix $\mathbf{W}$, $\mathbf{U}$ and $\mathbf{V}$ are the left and right singular vectors of $\mathbf{W}$, respectively, and $\tilde{\mathbf{\Sigma}}$ is a diagonal matrix containing the corresponding modified singular values,
\begin {align}
\tilde{\Sigma}_{ii}= \begin{cases}
	\Sigma_{ii} &\quad\text{if } \Sigma_{ii} < c\\
	c &\quad\text{if }\Sigma_{ii} \geq c.
\end{cases}
\end{align}
In practice, we set the upper bound on $c$ to be $0.9$ for numerical stability.

A hyperbolic tangent wrapper was applied to the output of the inverse pass of the iResNet (input of the forward pass), $\mathbf{x}_{K} \rightarrow \tanh(\mathbf{x}_{K})$, which maps all dimensions to a domain of $[-1,1]$. The final output of the inverse pass was linearly scaled from the range $[-1,1]$ to match the domain of the original vibrational coordinates:
\begin{equation}
	\label{eq:NF}
	f_\theta^{-1}(\mathbf{x}) = \mathbf{a}\cdot \tilde{f}^{-1}(\mathbf{x}) + \mathbf{b}.
\end{equation}
This ensures that $f_\theta^{-1}(\mathbf{x})$ is contained within the original coordinate domain for all possible values of the parameters $\theta$. 

\subsection{Loss function}
To optimize the parameters $\theta$ of $f_\theta$, we take advantage of the variational principle to define the loss function as
\begin{align} \label{eq:L_theta}
	\mathcal{L}_\theta^{M} = \sum_{n< M}E_n \rightarrow \min_\theta.
\end{align}

This loss function adapts the coordinates to the truncated basis, the chosen number of target states, and the specific molecule - each of which influences the predicted energies. When the exact eigenbasis is contained within the span of the truncated basis, the predicted energies are exact, and the loss reaches its minimum. 
If the number of target states is equal to the number of basis functions, $M=N$, this loss function can be efficiently computed as $\mathcal{L}_\theta^{N} = \Tr(\mathbf{H})$, eliminating the need to calculate off-diagonal matrix elements of the Hamiltonian matrix. Despite the added complexity, the computational cost of calculating off-diagonal matrix elements and repeated Hamiltonian matrix diagonalization can be justified when the optimization focuses on a specific subset of states of interest. 
In such cases, focusing on a smaller set of eigenvalues and selectively improving their accuracy can lead to a more efficient and targeted optimization process.

The loss function was optimized using the \texttt{Optax}~\cite{Optax:DeepMindJax} Adam optimizer with a learning rate of $0.001$, $\beta_1=0.9$, $\beta_2=0.999$, $\epsilon=10^{-8}$, and $\bar{\epsilon}=0.0$.

\subsection{Details of multidimensional calculations}
For the multidimensional calculations on H$_2$S and H$_2$O, the reference vibrational coordinates were chosen as conventional displacement-based valence coordinates, \ie, bond lengths and angles. Multidimensional basis functions were expressed as direct products of Hermite basis functions in all examples. The normalizing-flows architecture enables mapping of any input coordinate range, defined by the domain of $\textbf{r}$ (initial valence coordinates), to any output coordinate range, defined by the domain of $\textbf{q}$ (optimized coordinates). Therefore, Hermite basis functions are suitable for both the bond stretching coordinates, $(-\infty,\infty)\rightarrow [0,\infty)$, and the angular coordinates,  $(-\infty,\infty)\rightarrow [0,\pi]$. 

The direct-product basis was truncated by including only basis-product configurations $(n_1, n_2, n_3)$ that satisfy the polyad condition $2n_{1} + 2n_{2} + n_{3} \leq P_\text{max}$, where $n_1$, $n_2$, and $n_3$ represent the Hermite basis function indices corresponding to the two stretching and one bending valence coordinate, respectively. Two direct-product quadrature grids were used in an alternating fashion during training, with 30/31 points along the stretching coordinates and 60/61 points along the bending coordinate. These grids were chosen to minimize numerical integration error in the Hamiltonian matrix elements, keeping the optimization stable while maintaining computational efficiency. For higher-dimensional systems, it may be necessary to use sparse quadrature methods or Monte Carlo integration.

The PES used for H$_2$S and its isotopologues was obtained from~\citet{Azzam:MNRAS460:4063}, while the PES for H$_2$O was taken from~\citet{Conway:atmschemphys20:10015}.

\section{Results \& Discussion}

\subsection{Solving the Morse oscillator with harmonic oscillator basis functions}\label{sec:1dexample}

To better understand the morphology and effectiveness of the normalizing-flows coordinate optimization, we examine its application to a typical one-dimensional example of molecular vibrations: the Morse oscillator~\cite{Morse:PR34:57}. The Hamiltonian for the Morse oscillator is
\begin{align}
	H_M(r) = \frac{-\hbar^2}{2\mu}\frac{\partial^2}{\partial r^2} + D_e\left[1 - \exp(-a_M(r-r_e))\right]^2,
\end{align} 
where $\mu$ is the reduced mass of the oscillator, $D_e$ is the dissociation
energy, $a_M$ is a second Morse parameter, and $r-r_e$ is the displacement
coordinate relative to the  equilibrium bond distance, $r_e$. The nonlinear
coordinate transformation $z = 1 - \exp(-a_M(r-r_e))$ maps the Morse potential into a
harmonic potential,
\begin{align} 
	V_M(r) = D_e\left[1 - \exp(-a_M(r-r_e))\right]^2 \rightarrow \tilde{V}(z) = D_e z^2.
\end{align}
It may seem that the Hermite functions - eigenfunctions of the quantum harmonic oscillator - could be used to exactly solve the problem.
However, this coordinate transformation also modifies the kinetic energy operator. Specifically, the change in the volume element $ \mathrm{d}r\rightarrow  \mathrm{d}z$ results in the following kinetic energy operator expressed in the $z$ coordinate
\begin{align}
	\tilde{T}(z) = \frac{-\hbar^2}{2\mu}\left(a_M^2 (1-z)^2\frac{\partial^2}{\partial z^2} -a_M^2(1-z) \frac{\partial}{\partial z} \right).
\end{align}
This expression differs from the standard kinetic energy
operator for both the Morse and harmonic oscillators. It is therefore not
possible to map the Morse-oscillator Hamiltonian into the harmonic-oscillator
Hamiltonian by a change of coordinates.
This limitation can also be realized from the fact that a coordinate transformation does not change relative positions of nodes between different basis functions. The odd harmonic-oscillator eigenfunctions share a node at the center of their expansion, but the Morse oscillator eigenfunctions do not. In addition, the transformed coordinate $z$ is defined in the domain $(-\infty,1)$, which is incompatible with the domain $(-\infty,\infty)$ required for harmonic-oscillator eigenfunctions.

The optimal coordinate transformation can be determined by minimizing the loss
function in \eqref{eq:L_theta}. As demonstrated, this transformation does not
establish a one-to-one correspondence between the Hermite basis functions and
the Morse-oscillator eigenfunctions. Instead, the optimal transformation ensures
that the span of the $N$ Morse-oscillator eigenfunctions of interest,
$\{\Psi (r)\}_{n=0}^{N-1}$, is optimally captured within the span of the $M$
augmented harmonic-oscillator basis functions, $\big
\{\gamma\left(f_\theta(r); \theta\right)\big \}_{m=0}^{M-1}$.

We optimized linear and nonlinear coordinate transformations to obtain variational solutions of the Morse-oscillator problem. The potential parameters used were $a_M = 2.1440 \textup{~\AA}^{-1}$ ($1~\textup{\AA} = 10^{-10}$ m), $D_e = 42301\ \text{cm}^{-1}$, representative of a typical OH-stretching mode in molecules~\cite{Vogt:MP117:1629}.
Both coordinate transformations were optimized to approximate all 23 bound states using a truncated set of 23 Hermite functions. The normalizing-flows transformation was modelled by the iResNet with five residual blocks, which proved sufficient to achieve convergence of the coordinate.

\begin{figure}
	\includegraphics[width=\linewidth]{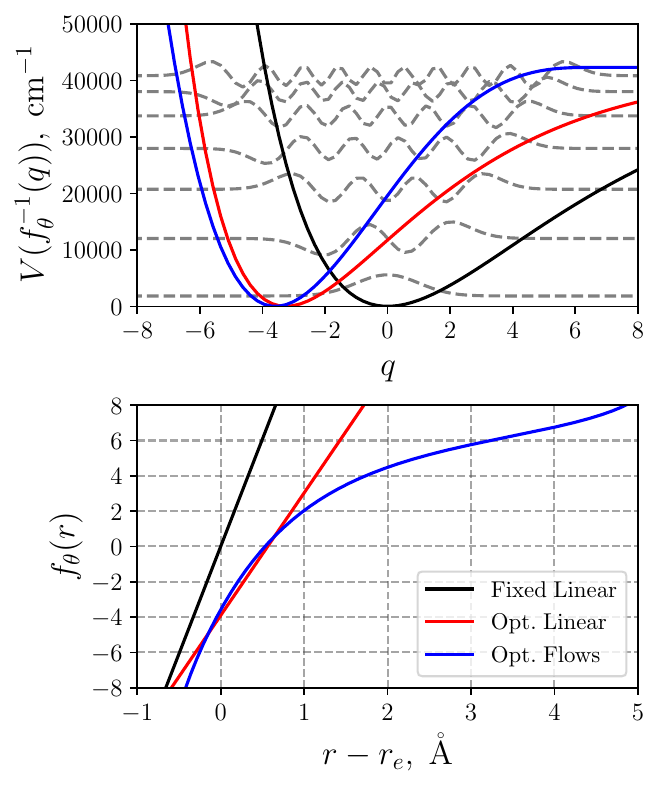}
	\caption{(Top panel) Morse potential expressed in fixed linear coordinates (black), optimized linear coordinates (red) and normalizing-flow coordinates (blue). The Hermite functions, $\mathcal{H}_n(q)$, with $n=0,3,...,18$ are depicted as grey dashed lines. (Bottom panel) The corresponding optimized coordinate $q = f(r;\theta)$ as a function of the displacement coordinate $r-r_e$.}
	\label{fig1}
\end{figure}

The Morse potential expressed in fixed linear (\textit{vide infra}), optimized linear, and normalizing-flow coordinates are shown in \autoref{fig1}.
In both of the optimized coordinates, the Morse potential is shifted relative to the Hermite basis reference at $r-r_e=0$, positioning the basis center closer to the average density center of the eigenstates, $\frac{1}{N}\sum_n \langle \Psi_n | r | \Psi_n \rangle$. This shift reflects that the optimal mapping is not a direct one-to-one correspondence between the Hermite basis functions and the Morse-oscillator eigenfunctions, $\gamma_n\rightarrow\Psi_n$. As a result, the largest contribution to the $n$-th eigenstate does not necessarily come from the $n$-th basis function. The largest difference between the linear and normalizing-flow coordinates is seen at large values of $r$ (approximately for $r-r_e>1 \ ${\AA}). In this region, the harmonic potential and the Morse potential differ significantly, leading to different decay behaviour of the respective eigenfunctions. 

The computed energy values for the Morse-oscillator problem expressed in fixed linear, optimized linear, and normalizing-flow coordinates are presented in \autoref{tab:morse1D}. The parameters $a$ and $b$ for the fixed linear coordinates are derived from a quadratic expansion of the potential in $r$ centered around $r_e$, followed by mapping of the resulting Hamiltonian onto the harmonic oscillator Hamiltonian in $q$. This yields,
\begin{align}\label{eq:fixed_params}
	a = \left(\frac{F }{\mu \hbar^2}\right)^{1/4} 
	,~~~ b=r_e,
\end{align}	
where $F =\frac{d^2 V(r)}{d r^2}\rvert_{r = r_e}$ is the force constant. The fixed linear transformation is effective for the lowest energy states, where wavefunctions remain largely confined within the region where the quadratic expansion of the potential holds. However, its accuracy declines rapidly with increasing level of excitation. In contrast, optimized linear coordinates perform worse for the lowest energy states but degrade more gradually in accuracy with increasing level of excitation. Normalizing-flow coordinates offer significantly improved accuracy across all energy levels, demonstrating the advantages of nonlinear coordinate transformations.

\begin{table}
	\vskip 0.15in
	\centering
	\small \vskip 5pt
	\begin{tabular*}{\linewidth}{@{\extracolsep{\fill}}lcccc}
		\toprule
		State &  Fixed linear & Opt. linear & Opt. flows & Reference  \\
		\midrule
		 0 & 0.00 & 0.09 & 0.00 & 1838.97 \\
		1 & 0.00 & 0.70 & 0.00 & 5394.32 \\
		2 & 0.00 & 6.16 & 0.00 & 8786.20 \\
		3 & 0.00 & 34.96 & 0.00 & 12014.61 \\
		4 & 0.01 & 99.80 & 0.00 & 15079.56 \\
		5 & 0.48 & 174.38 & 0.00 & 17981.04 \\
		6 & 9.76 & 263.49 & 0.00 & 20719.06 \\
		7 & 95.06 & 439.24 & 0.00 & 23293.61 \\
		8 & 458.58 & 712.07 & 0.00 & 25704.69 \\
		9 & 1312.18 & 1013.82 & 0.00 & 27952.31 \\
		10 & 2728.98 & 1279.12 & 0.01 & 30036.46 \\
		11 & 4729.23 & 1472.23 & 0.04 & 31957.14 \\
		12 & 7279.19 & 1583.27 & 0.11 & 33714.36 \\
		13 & 10473.30 & 1592.80 & 0.27 & 35308.11 \\
		14 & 14373.51 & 1651.39 & 0.53 & 36738.40 \\
		15 & 18484.27 & 1609.12 & 0.92 & 38005.22 \\
		16 & 24442.18 & 2375.75 & 1.47 & 39108.57 \\
		17 & 36614.07 & 3028.89 & 2.14 & 40048.46 \\
		18 & 58273.63 & 4412.65 & 2.78 & 40824.88 \\
		19 & 92645.97 & 6551.87 & 3.15 & 41437.83 \\
		20 & 145985.88 & 7988.28 & 3.65 & 41887.32 \\
		21 & 231296.45 & 13107.43 & 4.83 & 42173.35 \\
		22 & 382860.60 & 14762.55 & 11.53 & 42295.90 \\
		\midrule
		Sum & 1032063.33 & 64160.06 & 31.43 & 652300.37 \\
		\bottomrule
	\end{tabular*}
	\caption{Calculated bound state energy level discrepancies (in cm$^{-1}$) for the Morse potential using a basis of 23 Hermite functions. Results are provided for fixed linear, optimized linear, and normalizing-flows coordinate transformations. The reference energy levels were calculated analytically.}
	\label{tab:morse1D}
	\vskip -0.1in
\end{table}

The linear parameters in \eqref{eq:fixed_params} provide a well-established approach for modifying the coordinate mapping in response to changes in physical parameters of the system ($\mu$, $F$, and $r_e$).
However, the corresponding mapping strategy for the normalizing-flow coordinates is not entirely clear. To address this, in \autoref{sec:isotopologues} we investigate the transferability of the optimized normalizing-flow coordinates in isotopologues, \ie, molecules that share the same structure but differ in nuclear masses. 

\subsection{Transformations on finite and semi-finite domains: The need for nonlinear maps}\label{sec:trans_domains}
As illustrated in the previous subsection, the optimal coordinate ensures that the average density center of the basis functions is closely aligned with the average density center of the eigenfunctions. On infinite domains, this transformation can be conveniently accomplished using a simple linear transformation. However, on semi-finite and finite coordinate domains, linear transformations may produce unphysical results. 

Consider a one-dimensional Schrödinger equation defined on a domain $[r_\text{min}, r_\text{max}]$, where at least one of the limits is finite. This constrain commonly arises in radial and angular coordinates. Let $\left\{\phi_i\right\}_{i=0}^\infty$ be a basis set that is optimal for a given system on this domain, \eg, the eigenfunctions of the system. Suppose we wish to reuse this basis set for a different system defined on the same interval $[r_\text{min}, r_\text{max}]$, but with a shifted potential minimum and a different potential width. In principle, it is possible to account for such changes using a linear transformation of the form $L(r) = a r + b$, as suggested in \eqref{eq:fixed_params}. This transformation maps the original domain to a new domain, $L : [r_\text{min},r_\text{max}] \to [q_\text{min},q_\text{max}] = [L(r_\text{min}),L(r_\text{max})]$. Depending on the values of $a$ and $b$, the transformed domain may extend beyond the original physical region, i.e.,  $q_\text{min}<r_\text{min}$ and/or $q_\text{max}>r_\text{max}$, resulting in an unphysical domain. Even if the domain remains within bounds, the transformation may alter the implicit (Dirichlet or Neumann) boundary conditions of the basis functions, compromising their validity. 

Linear transformations on the infinite real line preserve the domain and avoid the aformentioned challenges on semi-finite or finite intervals. We exploit this property to construct coordinate maps between two different problems defined on the same (semi-)finite domain. We define a map $h: [r_\text{min}, r_\text{max}] \to [r_\text{min}, r_\text{max}], \quad r \mapsto q=h(r)$ as a composition of invertible transformations:
\begin{align}\label{eq:nonlinear}
	h = L_0^{-1} \circ T^{-1} \circ L \circ T \circ L_0,
\end{align} 
where $h = f\circ g$ denotes function composition $h(x)=f(g(x))$. The fixed linear transformation $L_0$ maps the initial domain $[r_\text{min},r_\text{max}]$ to the interval $I$, where $I$ is the domain of the nonlinear transformation $T : I \to (-\infty, \infty)$. After these two operations, the problem has been mapped into an infinite domain. Hence, any linear transformation $L$ preserves the domain. 
The inverse transformation, $T^{-1}$ followed by $L_0^{-1}$, returns the coordinate to the original physical domain  $[r_\text{min},r_\text{max}]$.
This construction defines a nonlinear map that contains an optimizable unconstrained linear transformation $L$, while remaining constrained to the physical domain.
We use the construction in \eqref{eq:nonlinear} in~\autoref{sec:molecules} as a fine-tuning mechanism for transfering coordinates from H$_2$S to H$_2$O. 

\begin{figure}[h!]
	\includegraphics[width=\linewidth]{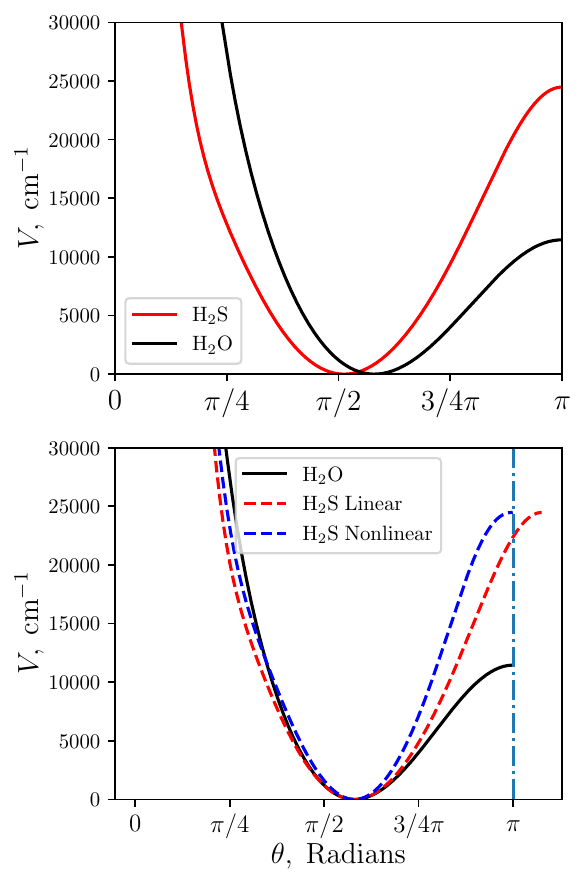}
	\caption{(Top panel) The one-dimensional bending PES of H$_2$S (red) and H$_2$O (black) with
	bond lengths fixed to the equilibrium values. (Bottom panel)
	A comparison between the H$_2$O	PES (black) and the composition of the H$_2$S with two different
	transformation optimised to try to approximate the water potential. The linear
	transformation $L^{-1}$ is shown in red and the nonlinear $h^{-1}$ in blue.
	The linear approach extends the potential beyond the physical domain, marked by the light-blue
	vertical line at $\theta=\pi$, while the nonlinear transformation keeps the potential within the
	domain.  
	}
	\label{fig2}
\end{figure}

To illustrate the limitations of linear coordinate transformations and the utility of the nonlinear transformation $h$, we consider a one-dimensional example defined on a finite interval. We construct two one-dimensional model potentials by fixing the valence radial coordinates at their equilibrium values and treating the valence angular coordinate as the variable. The systems considered are H$_2$S and H$_2$O, with the valence angular coordinate defined on the domain $[0, \pi]$. We treat the angular coordinate for H$_2$S as the one optimized for a chosen basis set and aim to reuse this basis and coordinate definition for H$_2$O. Rather than transforming the basis functions directly, which is difficult to visualize due to their infinite extent, we instead transform the coordinate in which the potential is expressed. This approach is analogous to transforming Hermite functions so that the second-order Taylor series of the potential at the minimum maps to that of the harmonic oscillator, as in~\eqref{eq:fixed_params}. The two potentials are illustrated in the top panel of \autoref{fig2}, which reveals differences in both equilibrium positions and potential shapes. 

We apply two different invertible transformations to the H$_2$S potential, each optimized to approximate the H$_2$O potential.
That is, we seek
\begin{align}
	V_{H_2S} \circ g^{-1} \approx V_{H_2O},
\end{align}
where $g$ is either a linear transformation $L$ or the nonlinear transformation $h$ defined in~\eqref{eq:nonlinear}. Both transformations are parametrized by only two adjustable parameters. The wrapper $T$ is chosen as the inverse hyperbolic tangent function.

The results of the optimizations are presented in the bottom panel of \autoref{fig2} alongside the original H$_2$O potential as a function of the valence-bond angle, which serves as the objective function. Both $L$ and $h$ succesfully shift the minimum and adjust the width of the composed H$_2$S potentials to match that of H$_2$O. Although the linear transformation produces a closer visual match to the H$_2$O potential, it also maps part of the domain outside the physically valid interval due to the absence of boundary constraints on the linear parameters. Applying such a transformation in vibrational calculations for H$_2$O would produce accurate results for only few low-energy states. Higher-energy eigenstates would sample unphysical regions of the Hamiltonian, resulting in unreliable or incorrect predictions. 

This problem is analogous to known challenges in using Hermite basis functions for semi-finite domains, \eg, bond-stretching vibrations. While in practice, numerical stability is often retained due to the steep rise of the potential at short bond length, this is not guaranteed. In contrast, our use of wrapping functions, \ie, invertible functions that map finite intervals to infinite ones such as $T$, within the normalizing-flow framework allows for unrestricted coordinate transformations on infinite, semi-infinite, and finite intervals while remaining withing the valid domain.

Although this example illustrates the limitations of direct linear coordinate transformations, it neglects the effects of coordinate transformations on the kinetic energy operator. Since our goal is to minimize computed approximate energies, any coordinate transformation must account for both potential and kinetic contributions. In some applications, such as isotopologue calculations explored in~\autoref{sec:isotopologues}, the systems share the same PES but differ in their kinetic operators. In such cases, a transformation of the form~\eqref{eq:nonlinear} remains suitable for mapping between solutions.

\subsection{Intuition of the shape of normalizing-flow coordinates}
In this section, we solve the one-dimensional Schrödinger equation for a simplified double-well potential, commonly used to model the inversion vibration in ammonia (NH$_3$).
The Hamiltonian is defined as follows:
\begin{align}
	H = -\frac{\hbar^2}{2\mu}\frac{\partial^2}{\partial r^2} + \frac{k (r^2 - r_0^2)^2}{8r_0^2},
\end{align}
where $\mu = \frac{3m_Hm_N}{3m_H+m_N}$, $k= \frac{8V_0}{r_0^2}$, $r_0=0.3816$ {\AA} and $V_0=2028.6$ cm$^{-1}$ is the inversion barrier. 

We solve the one-dimensional Schrödinger equation using the ansatz introduced in~\eqref{eq:gamma_n}, optimizing the invertible transformation to minimize the energies of the first few states. We use as many basis set functions as there are target states and repeat the optimization for different numbers of states.

\begin{figure}[h!]
	\includegraphics[width=\linewidth]{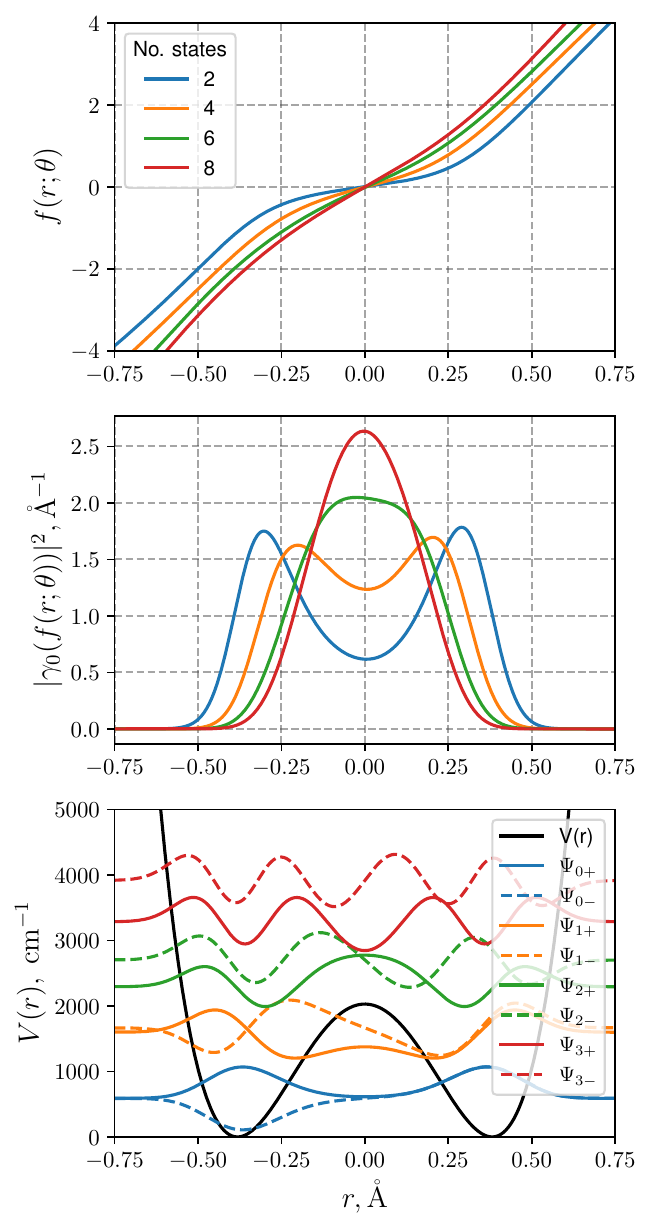}
	\caption{(Top panel) Optimized normalizing-flow coordinates for varying number of target states (using the same number of basis functions as target states in each case) as a function of the physical coordinate $r$. (Middle panel) The first augmented density distribution as a function of $r$. (Bottom panel) Potential energy curve along with the first few first associated eigenfunctions used in the coordinate optimization. Each eigenfunction is vertically shifted by its corresponding eigenvalue for visual purposes.
	}
	\label{fig3}
\end{figure}

In the top panel of \autoref{fig3}, we plot the optimized normalising-flow  coordinate transformations $q = f(r; \theta)$ obtained from these different optimizations.
When optimizing for only the two lowest states, the transformation develops a plateau around $r = 0$, where the gradient tends toward zero, and therefore so does the Jacobian determinant $D$. In the normalizing flow coordinate $q$, the plateau is centered around $q|_{r = 0} = f_\theta(0) = 0$, so the composed basis function $\gamma_0(q; \theta) = \phi_0(q)\sqrt{D}$, shown in the middle panel of \autoref{fig3}, becomes small near the origin. This occurs despite the primitive basis function $\phi_0(q) \propto \exp(-q^2/2)$ having a maximum at $q = 0$, and leads to a local minimum in $\gamma_0(q; \theta)$ at this point. As $r$ moves away from zero, the factor $\sqrt{D}$ increases faster than the exponential decay of the primitive basis function $\exp(-q^2/2)$, until the latter eventually dominates. This interplay between the gradient of the coordinate transformation and the shape of the primitive basis function creates a bimodal structure in the first composed basis function, starting from an originally unimodal Gaussian function. The first two eigenfunctions of the double-well potential are symmetric and antisymmetric in $r$, with near-zero and zero density at $r = 0$, respectively. This explains why the coordinate transformation creates a plateau in this region and why this improves the approximation power of the transformed basis compared to the original one.

As the number of target states increases, the mean density at $r=0$ across all eigenstates also increases (see the bottom panel of \autoref{fig3}). If the same transformation optimized for just the lowest two states were retained, all basis functions would have reduced amplitude around $r=0$, making it difficult to accurately represent higher-energy eigenstates, particularly those above the inversion barrier. For this reason, the optimized transformation becomes more linear, and the first composed basis function increasingly resembles a simple Gaussian.

Additionally, the top panel of \autoref{fig3} shows that the variation of the optimized coordinate transformation diminishes as the number of basis functions increases. For sufficiently large basis sets, the transformation exhibits minimal variation and stabilizes. This behaviour reflects the transferability of normalizing-flow coordinates across different basis-set truncations and helps to explain the ability of the method to consistently produce accurate vibrational energies.

\subsection{Transferability across basis-set truncations}\label{sec:trans_pmax}
During the training of the normalizing-flow mapping, the neural network leverages all available information from the truncated basis set to enhance its performance. As a consequence, the training disregards basis functions that are not included in the approximation, as exemplified in the previous subsection.
This might suggest that the resulting optimal coordinate mapping depend
strongly on the number of basis functions employed during training.
However, due to the inherent properties of normalizing flows, the transformed basis set retains completeness and orthogonality. As a result, the optimal coordinate mapping remains largely independent of the number of basis functions $M$, provided the basis set used for training is sufficiently expressive. 

The computational cost of training scales with the
number of basis functions, $M$, as $M^3$, if the loss function requires
solving an eigenvalue problem using full diagonalization techniques. It also depends on the complexity of the neural network.
The approximate independence from $M$ enables an efficient training protocol, introduced in Ref.~\citenum{Saleh:JCTC21:5221}, in which the normalizing-flow coordinates are first optimized using a small or moderate number of basis functions. These optimized coordinates can then be transferred to
calculations with larger basis sets for higher accuracy in energy predictions. 
We refer the reader to Ref.~\citenum{Saleh:JCTC21:5221} for more details on this
transferability approach. In \autoref{sec:isotopologues} and  \autoref{sec:molecules},
we make use of this property to reduce computational cost while maintaining accuracy.

\subsection{Transferability across isotopologues}\label{sec:isotopologues}
In this section, we investigate the effectiveness of transferring normalizing-flow coordinates between isotopologues of H$_2$S. The coordinates were initially optimized in variational calculations for the lowest 100 vibrational energy levels of H$_2$S using a basis set truncated at $P_\text{max}=12$, corresponding to 140 basis functions. The optimized coordinates were then applied in calculations of the vibrational energy levels of D$_2$S and HDS.  Non-Born-Oppenheimer effects are neglected, and the PES is therefore approximated to be invariant across isotopologues.

\begin{figure}[h!]
	\includegraphics[width=\linewidth]{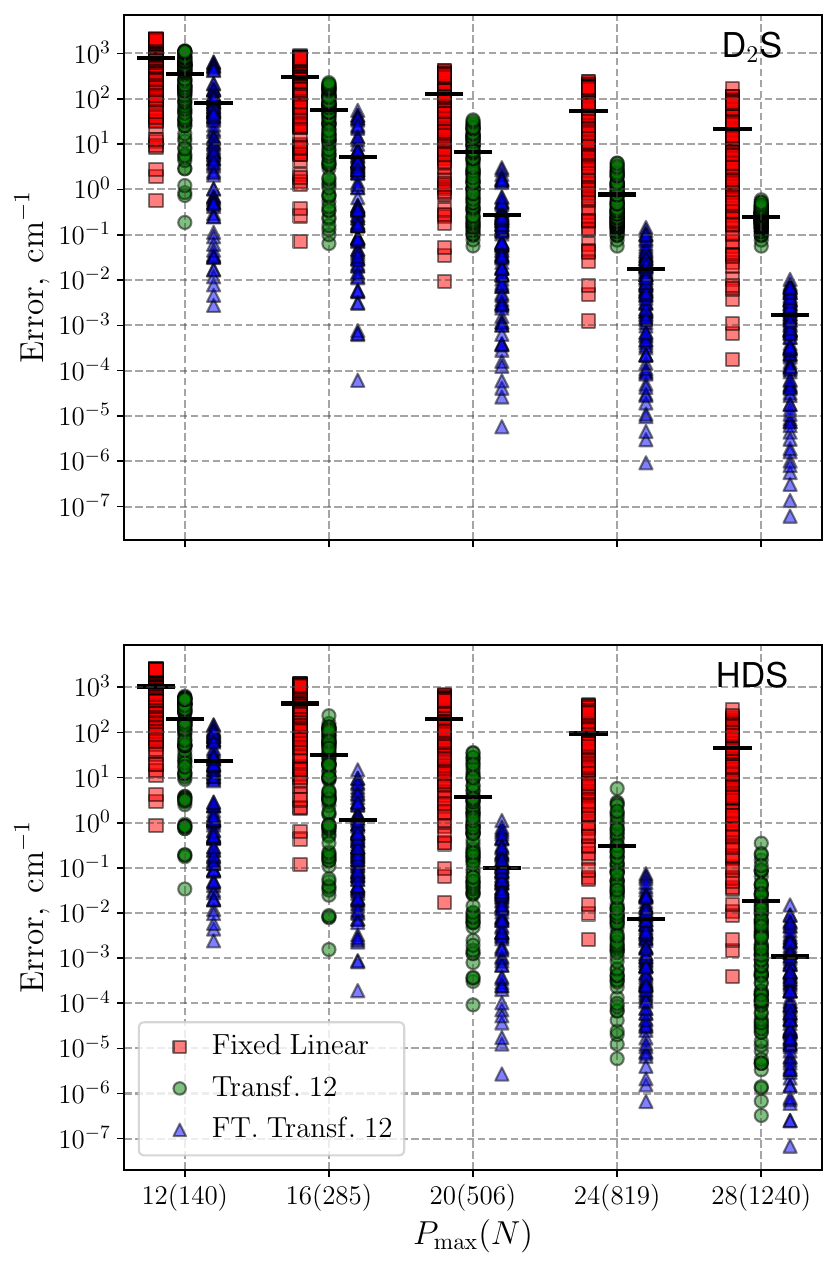}
	\caption{Convergence of the first 100 vibrational energy levels of (a) D$_2$S and (b) HDS using fixed linear transformation of valence coordinates (red squares), transferred normalizing-flow coordinates (green circles),
	and fine-tuned transferred normalizing-flow coordinates (blue  triangles). The vertical axis represents the calculated
	energy error, $E_i - E_i^{(\text{Ref})}$, while the horizontal axis shows
	the basis-set truncations, $P_\text{max}$. The transferred
	coordinates were originally optimized for H$_2$S with $P_\text{max}=12$, and the
	fine-tuned parameters were obtained through re-optimization of the linear parameters in the
	normalizing-flows model. The black horizontal lines show
	the average error per energy level. The data points are slightly offset along the $P_\text{max}$ axis for visual clarity.}
	\label{fig4}
\end{figure}

In \autoref{fig4}, we show a direct comparison of vibrational energies computed using the transferred normalizing-flow coordinates with those obtained using a fixed linear transformation of valence coordinates. In contrast to the transferred normalizing-flow coordinates, the fixed linear transformation inherently accounts for mass changes when applied to different isotopologues. Overall, the transferred normalizing-flow coordinates yield more accurate energy levels compared to the valence coordinates, with error reductions of up to three orders of magnitude. Additionally, they show faster convergence with respect to the basis-set truncation $P_\text{max}$. This improvement in accuracy is
more pronounced for HDS, particularly for lower vibrational levels,
which are generally better converged.
We attribute this improved accuracy to the enhanced separability of the Hamiltonian when expressed in normalizing-flow coordinates~\cite{Saleh:JCTC21:5221}.

Enhanced separability means that the vibrational energy levels can be more accurately approximated as sums of contributions from individual vibrational modes. However, isotopic substitution affects this separability. For example, when transitioning from H$_2$S to
HDS, the Hamiltonian separability is slightly reduced.
In such cases, the transferred flow coordinates can still provide reasonably accurate predictions for low-energy states, although accuracy diminishes for higher vibrational excitations.
When both hydrogen atoms are replaced, as in the transition from H$_2$S to D$_2$S, the Hamiltonian becomes even less separable.
As a result, the transferred coordinates are less effective, and the 
accuracy of energy predictions declines, even for the low-energy vibrational states. 

To further improve the performance of the transferred flow coordinates, we fine-tuned the linear model parameters $\mathbf{a}$ and $\mathbf{b}$ in \eqref{eq:NF}. Due to their greater mass, the wavefunctions of the heavier isotopologues are more localized, thus allowing optimization of the linear parameters without risk of mapping outside the physical domain. This eliminates the need for the more complex nonlinear optimization described in \autoref{sec:trans_domains}. For D$_2$S and HDS, the fine-tuning parameters we obtained by solving the Schrödinger equation for 100 vibrational states, using just 40 iterations with the Adam optimizer. The fine-tuning was performed only for the smallest truncated basis set, $P_\text{max}=12$. The resulting optimized coordinates were subsequently transferred to calculations with larger truncated basis sets, without reoptimization of the parameters, as described in \autoref{sec:trans_pmax}. The results for the fine-tuned coordinates in~\autoref{fig4} show a significant improvement of computed energy levels, by up to more than three orders of magnitude, compared to the results obtained using a fixed linear transformation of valence coordinates.

The rationale for fine-tuning transferred coordinates across isotopologues is well founded, as the fixed linear mapping of valence coordinates already includes mass-dependent scaling (see \eqref{eq:fixed_params}). Based on this, the variation in the linear parameters of the normalizing-flow model is expected to scale proportionally to the diagonal elements of the mass-weighted metric tensor, \ie, $a_i\propto
G_{ii}^{1/4}$ and  $b_i\rightarrow0$ as $G_{ii}\rightarrow 0$, consistent with the eigenstate average density center argument discussed in \autoref{sec:1dexample}. For valence-stretching vibrations, $G_{ii} = \mu^{-1}$, as used in \eqref{eq:fixed_params}.

\subsection{Transferability across molecules}\label{sec:molecules}
Conventional geometrically defined curvilinear coordinates are broadly applicable across various molecules. For example, valence coordinates are particularly well suited for semi-rigid molecules like H$_2$S, H$_2$O and H$_2$CO, and can be conveniently expressed using a Z-matrix representation.
In the previous subsection, we demonstrated how optimized normalizing-flow coordinates can be effectively transferred across isotopologues.
Here, we extend this analysis to evaluate the performance of normalizing-flows coordinates when transferred between different molecules with similar chemical structures.

\begin{figure}[h!]
	\includegraphics[width=\linewidth]{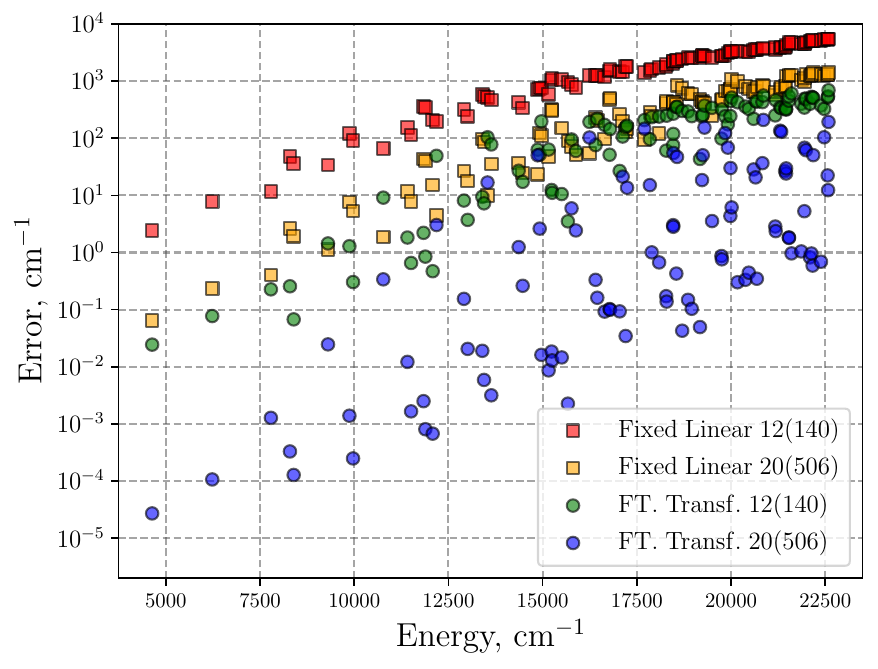}
	\caption{Convergence of the first 100 energy levels of H$_2$O for $P_\text{max}=12$ (140 basis functions) and $P_\text{max}=20$ (506) using fixed linear transformed valence coordinates (red and orange squares) compared to fine-tuned (FT.) normalizing-flow coordinates transferred from H$_2$S calculations, optimized for $P_\text{max}=12$, for $P_\text{max}=12$ and $P_\text{max}=20$ (green and blue circles). The vertical axis represents the calculated
		energy error, $E_i - E_i^{(\text{Ref})}$, while the horizontal axis shows
		the reference energy, $E_i^{(\text{Ref})}$. The converged energy values for H$_2$O were computed using optimized normalizing-flow coordinates with $P_\text{max}=24$.}
	\label{fig5}
\end{figure}

In \autoref{fig5}, we compare the convergence of the first 100 vibrational energy levels of H$_2$O computed with basis sets truncated at $P_\text{max}=12$ and $P_\text{max}=20$, using two different coordinate systems: (1) a fixed linear transformation of valence coordinates, and (2) normalizing-flow coordinates optimized for H$_2$S with $P_\text{max}=12$. The normalizing-flow coordinates were fine-tuned using the transformation discussed in \autoref{sec:trans_domains}, with six parameters, one multiplicative and one additive parameter for each coordinate. The fine-tuning was performed only for the smallest truncated basis set, $P_\text{max}=12$, and subsequently transferred without reoptimization to the larger truncation $P_\text{max}=20$ (see \autoref{sec:trans_pmax}).
The results in \autoref{fig5} show that normalizing-flow coordinates transferred from H$_2$S to H$_2$O yield more accurate predictions for all 100 vibrational energy levels compared to the valence coordinates, for the same $P_\text{max}$.
This demonstrates the utility of knowledge transfer between the Hamiltonians of chemically related molecules H$_2$S and H$_2$O. The observed improvement is expected, given the morphological similarity of their PESs, which also underlies the general effectiveness of valence coordinates in both systems.

While H$_2$S and H$_2$O exhibit similar structures in their lowest-energy vibrational states, this resemblance breaks down at higher excitations of the bending mode. Notably, H$_2$O begins to sample quasilinear configurations within the first 100 vibrational states, unlike H$_2$S. These are exactly the states in \autoref{fig5} for which the convergence deteriorates when using transferred normalizing-flow coordinates. Despite this important difference in high-energy behaviour, the normalizing-flow coordinates optimized for H$_2$S still provide a superior representation of the bending motion in H$_2$O compared to valence coordiantes, within the same truncated basis set. 

\section{Conclusion}
The normalizing-flow approach offers a versatile framework for optimizing vibrational coordinates in variational calculations of molecular vibrational energies.
In previous work, we showed that optimized normalizing-flow coordinates produce physically meaningful transformations that enhance the separability of the Hamiltonian when expressed in a direct-product basis~\cite{Saleh:JCTC21:5221}.

A key question is whether the normalizing-flow method can learn intrinsic
coordinates that not only improve computational performance for individual
molecules, but that also generalize across chemically related systems.
In this work, we took an initial step toward
addressing this question.

We demonstrated that optimizing vibrational coordinates using normalizing flows can improve the accuracy of variationally computed vibrational energies by several orders of magnitude. Furthermore, these coordinates enable much faster basis-set convergence compared with both fixed linear and optimized linear transformations of valence coordinates.

Although the optimized coordinates are tailored to a specific molecule, basis-set truncation, and target set of eigenstates, our results show that this specificity does not preclude generalization. On the contrary, we find that the same coordinates often remain effective across different basis-set truncations, isotopologues, and even across molecules with similar structural motifs.

This findings suggest that normalizing flows may
uncover coordinate systems that
reflect fundamental features of molecular motion. While these coordinates are
optimized for spectral accuracy, their capacity to generalize indicates a deeper
connection to the intrinsic vibrational coordinates long sought in theoretical
chemistry. Thus, variational coordinate optimization with normalizing flows
may serve not only as a practical computational tool, but also as means of generating physically motivated representations of vibrational motion.

Looking ahead, several promising directions emerge. One is to develop a generalized, trainable coordinate mappings that embed molecular descriptors or parameters directly into the model, thereby enabling automatic adaptation across different chemical systems. Another is to extend the current framework to larger molecular systems. At present, the use of dense Gaussian quadrature grids imposes a practical limit on system size. To address this, we are investigating Monte Carlo integration as an alternative. While it introduces greater integration errors, it offers a viable path to higher-dimensional systems where traditional quadratures become unfeasible.

Finally, our current architecture does not yet incorporate molecular symmetry, which is known to improve efficiency in conventional vibrational treatments~\cite{Yurchenko:JCTC13:4368}. Enforcing symmetry in the coordinate transformation and basis function could provide similar benefits for normalizing-flow coordinates. However, imposing symmetry constraints may reduce the flexibility of the neural network, presenting a tradeoff between generalization and effectiveness.

\section*{Acknowledgments}
The authors would like to acknowledge Jochen Küpper and the CFEL Controlled Molecule Imaging group for scientific discussion and organisational support.
The authors also acknowledge Henrik G. Kjaergaard for useful comments on the manuscript.

This work was supported by Deutsches Elektronen-Synchtrotron DESY, a member of the Helmholtz Association (HGF), including the Maxwell computational resource operated at DESY, by the Data Science in Hamburg HELMHOLTZ Graduate School for the Structure of Matter (DASHH, HIDSS-0002), and by the Deutsche Forschungsgemeinschaft (DFG) through the cluster of excellence ``Advanced Imaging of Matter'' (AIM, EXC~2056, ID~390715994). This project has received funding from the European Union's Horizon Europe research and innovation programme under the Marie Skłodowska-Curie grant agreement No. 101155136.

\section*{Author contributions}
All authors conceptualized the work, developed the theory, produced the code, and wrote the manuscript. E.V, A.F.C, and Y.S. conducted the calculations and analysed the results.

\section*{Notes}
The authors declare no competing financial interest. \\
Published open access in The Journal of Chemical Physics, 2025, 163 (15), 154106. Available at: \url{https://doi.org/10.1063/5.0285954}

\section*{Data availability}
The data that support the findings of this study are available within the
article and through the repository:
\url{https://gitlab.desy.de/CMI/CMI-public/flows/-/releases/v.0.2.1}. 

\section*{Code availability}
The code utilized and further developed in this work is publicly available at \url{https://gitlab.desy.de/CMI/CMI-public/flows/-/releases/v.0.2.1}.

\bibliography{string,cmi}%
\onecolumngrid%
\listofnotes%
\end{document}